\documentclass[runningheads]{llncs}
\usepackage[T1]{fontenc}
\usepackage{graphicx}
\usepackage{siunitx}
\usepackage{tabularray}
\usepackage{booktabs}
\usepackage{subfigure}
\usepackage{multirow}
\usepackage[normalem]{ulem}
\usepackage{hhline}

\usepackage{amssymb}
  \usepackage{graphicx}
  \usepackage{amsmath}
  \usepackage{siunitx}
  \usepackage{bbding}

\begin{document}
\title{Designing a Meta-Reflective Dashboard for Instructor Insight into Student–AI Interactions}
\titlerunning{Meta-Reflective Dashboard for Instructor Insight of Student–AI Interactions}

\author{
Boxuan Ma\inst{1}
\and
Baofeng Ren\inst{1}
\and
Huiyong Li\inst{1}
\and
Gen Li\inst{1}
\and
Li Chen\inst{2}
\and \\
Atsushi Shimada\inst{1}
\and
Shin'ichi Konomi\inst{1}}
\authorrunning{Ma et al.}

\institute{Kyushu University, Fukuoka, Japan  \\
\and
Osaka Kyoiku University, Osaka, Japan
}


%
\maketitle              % typeset the header of the contribution
\begin{abstract}

Generative AI tools are increasingly used for coursework help, shifting much of students’ help-seeking and reasoning into student–AI chats that are largely invisible to instructors. This loss of visibility can weaken instructors’ ability to understand students’ difficulties, ensure alignment with course goals, and uphold course policies. Yet transcript-level access is neither scalable nor ethically straightforward: reading raw chat logs across a class is impractical, and exposing detailed dialogue can raise privacy concerns and chilling effects on help seeking. As a result, instructors face a tension between needing actionable insight and avoiding default surveillance of student conversations. To address this gap, we propose a meta-reflective dashboard that makes student–AI sessions interpretable without exposing raw chat logs by default. After each help-seeking session, a reflection AI produces a structured, session-level summary of the student’s interaction trajectory, AI usage patterns, and potential risks. We co-designed the dashboard with instructors and students to surface key challenges and design goals, and conducted a formative evaluation of perceived usefulness, trust in the summaries, and privacy acceptability. Findings suggest that the proposed dashboard can reduce instructors’ sensemaking effort while mitigating privacy concerns associated with transcript-level access, and they also yield design implications for evidence, governance, and scalable class-level analytics for AI-supported learning.

\end{abstract}
\section{Introduction}

Learning analytics dashboards that track and visualize student activity play an important role in helping instructors understand students’ engagement and learning progress, drawing on data such as page views, quiz scores, and assignment submissions collected through conventional Learning Management Systems (LMS) \cite{jivet2018license,verbert2020learning}. Recently, many learning platforms have begun integrating chat-based AI tools \cite{shimada2025leveraging}. These tools provide timely support and lower the barrier to asking questions, but they also shift much of the help-seeking and problem-solving process into private chat interactions that current dashboards cannot capture \cite{ortega2025configuring}. As a result, instructors often cannot see what students asked, what explanations or answers were provided, whether these align with course goals, or whether submitted work is largely based on AI-generated responses \cite{ortega2025configuring}.

A central challenge in this context concerns teacher autonomy, defined as teachers’ willingness, capacity, and freedom to take control of their own teaching and learning \cite{vangrieken2017teacher}. Supporting teacher autonomy in the GenAI era is critical, as instructors contribute contextual and relational expertise, including socio-emotional support and course-specific pedagogical intent (e.g., learning objectives and assessment expectations), that should guide how AI is integrated and governed in classroom settings\cite{giannakos2025promise,ortega2025configuring}. When students use AI for coursework, instructors may be unaware of students’ prompts and the AI’s responses, whether these interactions align with course objectives, or the extent to which submitted work reflects copying \cite{ortega2025configuring}. As a result, limited visibility into student–AI interactions can undermine teacher autonomy and reduce instructional effectiveness.

Survey evidence suggests that many students self-report AI-enabled cheating \cite{lee2024cheating}, while instructors also express substantial concern about AI’s implications for academic integrity \cite{jones2020we}. Prior work has also shown that many instructors express a need for monitoring dashboards that reveal how students use AI tools, as such information can support integrity oversight and highlight common misconceptions and gaps in instruction \cite{kazemitabaar2024codeaid}. At the same time, accessing individual-level interaction data raises privacy concerns: even when intentions are pedagogical, students may feel surveilled to some extent and become less willing to ask questions \cite{ifenthaler2016student,ortega2025configuring,rubel2016student}. In addition, reading full chat logs for many students is unrealistic at scale, and it remains difficult to quickly distinguish when AI is supporting productive learning and when it is replacing meaningful thinking. 

To address these tensions, we transform student–AI chats into privacy-aware, session-level signals through a teacher-facing meta-reflective dashboard that distills each student–AI session into a concise, high-level summary of the student’s help-seeking and thinking process. Instead of exposing raw chat logs, the system generates session-level meta-reflective overviews that highlight interaction patterns and potential issues, providing support for instructor sensemaking and enabling scalable class-level analytics. We conduct a co-design process with instructors to identify key design goals and privacy boundaries, and then evaluate a prototype with instructors and students to examine interpretability, perceived usefulness, trust, and privacy acceptability.

This paper makes three contributions. First, we introduce a privacy-aware meta-reflective dashboard that generates structured, session-level summaries capturing interaction patterns and potential risks, enabling rapid instructor sensemaking without transcript-level access. Second, we report design goals and privacy boundaries identified with instructors and students. Third, we present formative evaluation findings from instructors and students on interpretability, perceived usefulness, trust, and privacy acceptability. Our study provides design insights for future instructor dashboards that aim to make student–AI interactions interpretable at scale while respecting privacy constraints.

%Although instructors are the primary audience, the meta-reflective summaries may also benefit students and AI tutors. For students, the summaries can model how to review their own interactions, formulate more effective prompts, and adopt stronger metacognitive and problem-solving strategies. For AI tutors, aggregating these reflections can reveal how the system behaves in authentic learning scenarios and where its guidance may be unhelpful, superficial, or misaligned with course goals, offering actionable directions for improving prompting policies and tutor behavior.

\section{Related Work}

\subsection{AI in Education}

GenAI tools, such as ChatGPT, have emerged as transformative technologies in higher education. These tools provide automated feedback, generate solutions, and enhance learning experiences, reshaping how students seek help and make progress on learning tasks \cite{holmes2022state,qu2025generative}. Compared with conventional help channels such as office hours, chat-based AI lowers the barrier to asking questions, enables rapid iteration, and offers on-demand explanations and examples \cite{zhang2021ai}. In practice, this can broaden participation in help seeking and support students when instructors are unavailable.

However, AI-supported learning introduces new risks and tensions. The same convenience that makes help seeking easier can also encourage over-reliance and superficial progress, and may increase the risk of cheating, when students focus on obtaining answers rather than building understanding \cite{ma2024enhancing,fan2025beware}. Prior studies suggest that excessive dependence on AI may undermine the development of critical thinking and independent learning \cite{lyu2024evaluating}. Even when GenAI tools streamline help seeking, they may discourage deeper cognitive engagement if students are not encouraged to critically evaluate and refine AI-generated outputs \cite{ma2024enhancing}.

More importantly, because much of the reasoning and decision-making happens in student--AI conversations that are often not visible to instructors, learning-relevant processes become less observable through traditional course artifacts. As a result, instructors may have difficulty understanding what students struggled with, how they used AI support, and whether these interactions align with course goals and assessment expectations \cite{ortega2025configuring}. These shifts are not only pedagogical but also affect instructional oversight: when learning processes move into student--AI chats that instructors cannot readily see, instructors may lose the actionable visibility needed to steer instruction and enforce course policies. This directly raises questions about teacher autonomy in AI-mediated learning.

\subsection{Teacher Autonomy}

Teacher autonomy refers to teachers' willingness, capacity, and freedom to take control of their own teaching and learning \cite{huang2005teacher}. In instructional settings, this autonomy is reflected in teachers' perceived decision authority over core domains such as assessment, teaching methods, and curriculum enactment.

The Human--AI Automation Model conceptualizes instructional autonomy as six levels of control, ranging from contexts where teachers retain full autonomy to those where technology assumes full autonomy \cite{molenaar2022towards}. Between these extremes, instructors may monitor student behavior, be alerted when attention is needed, or select, approve, and shape system reactions. Viewed through this lens, GenAI-enabled learning environments risk shifting control and informational advantage away from instructors, especially when learning-relevant interactions occur in student--AI chats that instructors cannot easily observe or interpret.

Maintaining teacher autonomy is especially important in GenAI contexts because high-quality teaching depends on forms of expertise that current AI systems do not reliably capture, such as pedagogical judgment, alignment with course objectives, and sensitivity to learners' affect and motivation \cite{giannakos2025promise}. When instructors must adapt to opaque AI behaviors or when actionable information about students' AI-mediated help seeking is unavailable, their autonomy can be weakened in practice. Accordingly, prior work calls for tighter human--AI synergy so that AI support remains aligned with learning goals and instructors can stay agentic when monitoring, evaluating, and shaping AI-mediated learning processes \cite{chan2024will,giannakos2025promise}.

\subsection{Learning Analytics Ethics}

While teacher autonomy motivates increasing instructors' visibility into student--AI help seeking, such visibility must be balanced against learning analytics ethics, privacy, and student trust \cite{sun2019s,jivet2018license}. A central design question is how much detail is necessary and justified for teaching purposes: finer-grained traces may improve interpretability, yet they can also increase surveillance risk and the likelihood of misuse \cite{sun2019s,korir2023investigating}.

Prior work shows that students' acceptance of learning analytics depends on perceived legitimacy, clear pedagogical purposes, and meaningful privacy protections. Discomfort rises when learners feel continuously tracked, even when tracking is framed as support \cite{jones2020we,ifenthaler2016student,korir2023investigating}. These concerns become more acute for AI-mediated learning because chat transcripts may include sensitive personal information, off-task content, or exploratory questions that students would not share in graded artifacts. Accordingly, ethical analytics for student--GenAI interactions should provide actionable and interpretable signals while avoiding default access to raw dialogue content \cite{ifenthaler2016student,jones2020we,ortega2025configuring}. This motivates our privacy boundary: rather than enabling transcript-level surveillance, we emphasize concise, session-level indicators that support pedagogical action while reducing privacy intrusion and potential chilling effects on help seeking.

Taken together, prior work suggests a tension between instructors’ need for actionable insight into student–AI help seeking and the ethical and practical limits of transcript-level monitoring. This motivates our meta-reflective dashboard approach, which summarizes each student–AI session into concise, privacy-aware signals to support instructor sensemaking and timely follow-up.

\section{Designing the Meta-Reflective Dashboard}

\subsection{System Development Approach}

The development of our teacher-facing dashboard follows a human-centered process. In this paper, we report an initial design-and-evaluation cycle. We first conducted conceptual co-design sessions with instructors (n=4) and students (n=2) to clarify classroom needs, acceptable privacy boundaries, and the types of session-level signals that would be meaningful for teaching. We then translated these insights into a set of design challenges and corresponding design goals, which guided the design of our system architecture and prototype. Finally, we conducted an initial evaluation with instructors (n=6) and students (n=8) to examine interpretability, perceived usefulness, trust, and privacy acceptability, and to identify directions for refinement in the next iteration.

\subsection{Design Challenges and Design Goals}

The co-design sessions surfaced stakeholder concerns that we translated into design challenges and goals. Building on the co-design discussions, we identified three challenges that instructors face when student learning processes shift into student--AI chat interactions: (1) \textbf{Privacy and surveillance concerns}, where accessing individual-level conversations may make students feel monitored and less willing to ask candid questions; (2) \textbf{Instructor workload}, which limits the feasibility of transcript-level monitoring because reading full chat logs across many students is impractical in everyday teaching; and (3) \textbf{Actionability}, as instructors may struggle to quickly distinguish productive AI use from over-reliance and to decide when intervention is warranted even when interaction data is available. These challenges informed the following design goals.

\textbf{[DG1] Provide privacy-aware student--AI interaction monitoring.}
To address privacy and surveillance concerns, the dashboard should make student--AI interactions interpretable through session-level, high-level representations by default, rather than exposing raw chat logs. It should support clear privacy boundaries so that instructors can gain insight while reducing the risk of normalizing transcript-level surveillance.

\textbf{[DG2] Enable high-level, actionable instructor sensemaking.}
To address instructor workload, the dashboard should distill each session into concise, structured, and interpretable summaries that enable rapid understanding without extensive reading. The presentation should prioritize clarity and low effort, helping instructors quickly identify what happened in a session and what may require attention.

\textbf{[DG3] Highlight risk signals and support timely instructor follow-up.}
To address actionability, the dashboard should highlight potential risk patterns (e.g., direct answer-seeking, lack of verification, repeated misconceptions) and pair them with lightweight, actionable follow-up suggestions. Risk signals should be framed as indicators rather than definitive judgments, supporting timely intervention while reducing the likelihood of overreaction or misuse.

In addition, to support adoption, the system should be designed to integrate with LMS via single sign-on, as in previous work \cite{ortega2025configuring}, so instructors can access it within their existing course workflow without registering for a separate tool. 

\begin{figure}[t]
\centering
\includegraphics[scale=0.35]{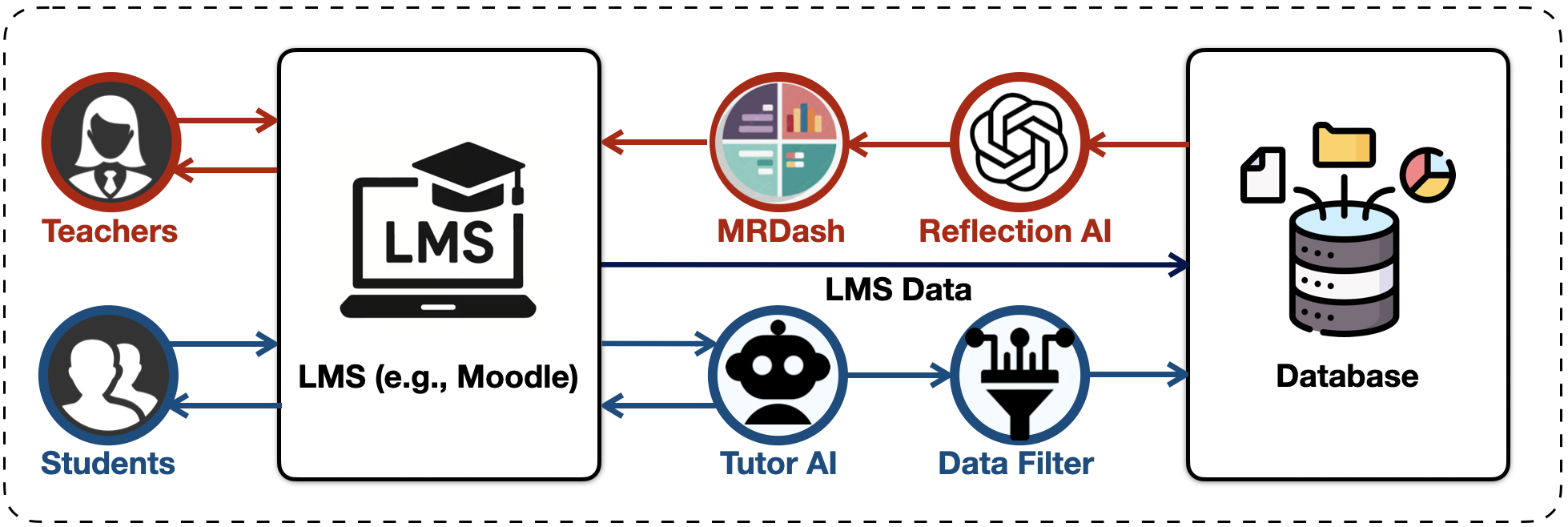}
\caption{Overview of the proposed system architecture.}
\label{archi}
\end{figure}

\subsection{System Architecture}

Figure~\ref{archi} illustrates the proposed system architecture. The system is designed to be embedded within existing LMS and accessed through single sign-on to reduce adoption friction and fit instructors’ everyday workflow. In practice, it can be deployed as an LTI-compatible tool, enabling role-based access for instructors and students while reusing course context and existing access control.

Our approach treats each student--AI help-seeking session as the basic unit of analysis and separates two roles: a \textit{Tutor AI}, which interacts with the student during the task, and a \textit{Reflection AI}, which analyzes the conversation and generates a meta-reflective report.

On the student side, learners interact with the \textit{Tutor AI} within the LMS during a help-seeking or problem-solving episode. Student prompts and AI responses are routed through a \textit{Data Filter} before being stored. The \textit{Data Filter} enforces privacy-aware logging (e.g., removing sensitive information and applying course-specific logging policies) so that the system can support instructional insight without exposing raw chat transcripts. The filtered data are then written to a \textit{Database} as the system’s learning record store.

After the session concludes, the dialogue record for that session is forwarded to the \textit{Reflection AI}. The \textit{Reflection AI} produces a structured meta-reflective report for the specific session. It organizes student prompts into different types and groups these types into higher-level categories. Based on the resulting distribution and the dialogue content, it infers key interaction patterns, highlights potential risks (e.g., copy \& paste behavior), and generates a short session summary. 

%The definitions of the higher-level categories and prompt types are informed by prior work \cite{ma2025scaffolding} and can be customized to fit different courses and tasks.

The resulting report is passed to the instructor-facing dashboard embedded in the LMS, supporting sensemaking and follow-up without requiring instructors to inspect full transcripts. In addition, the system can incorporate \textit{LMS data} (e.g., learning materials and assignment context) to situate each session within the surrounding learning activity and to support future extensions such as linking interaction patterns with student performance indicators.

\begin{figure}[t]
\centering
\includegraphics[width=\textwidth]{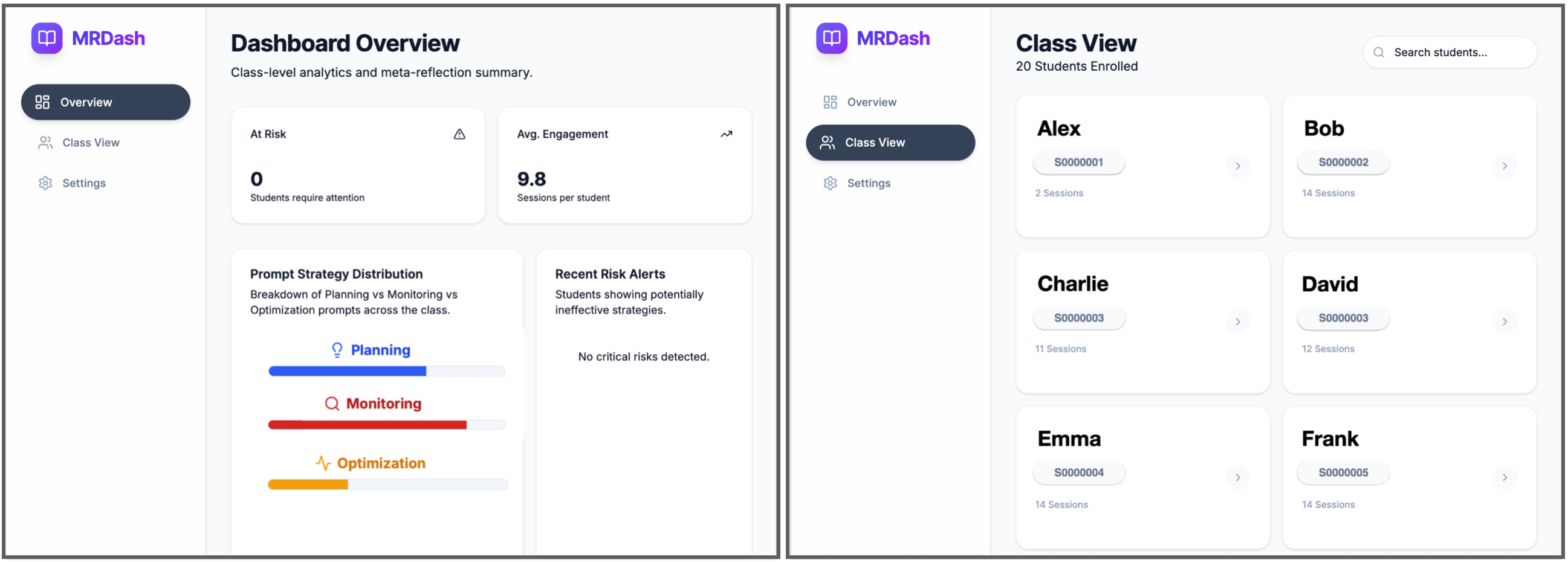}
\caption{Prototype interface of MRDash showing the class-level overview and class view.}
\label{overview}
\end{figure}

\subsection{System Prototype}

We implemented an initial prototype, \emph{MRDash}, to operationalize the three design goals. 

%The prototype was populated with real-world student-AI dialogue data from an introductory Python programming course, and all student identities were represented using pseudonymous identifiers.

\paragraph{Overview and Class View Interfaces.}

Figure~\ref{overview} shows two instructor-facing interfaces. The \textit{Overview} page provides a lightweight class snapshot, including an at-risk indicator and aggregate distributions of high-level prompt strategy categories (e.g., planning, monitoring, optimization). These summary statistics are intended to support quick situational awareness without requiring instructors to inspect individual transcripts. The \textit{Class View} page lists student names, student IDs and simple activity indicators (e.g., number of recorded sessions), enabling instructors to locate specific students or triage cases for follow-up.

\begin{figure}[t]
\centering
\includegraphics[width=\textwidth]{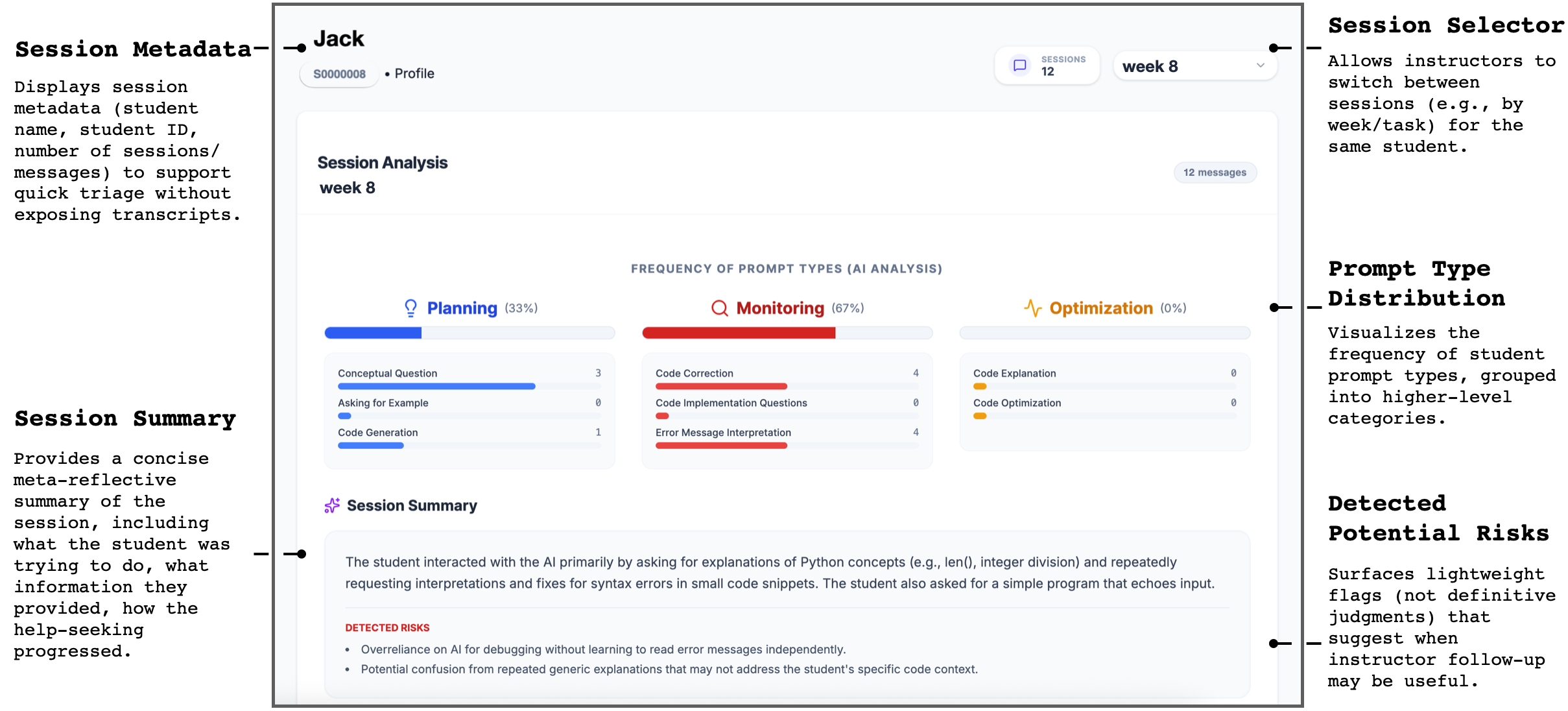}
\caption{Prototype interface of MRDash showing the session-level analysis view.}
\label{sessionview}
\end{figure}

\paragraph{Session-level Analysis Interface.}
Figure~\ref{sessionview} presents the \textit{Session View}, the core analytic unit in MRDash. For a selected student and help-seeking session (e.g., filtered by week or topic), the dashboard summarizes basic session context (e.g., number of messages) and visualizes the \textit{prompt-type distribution}. Student prompts are coded into fine-grained prompt types and grouped into higher-level categories using a taxonomy adapted from prior codebooks on metacognitive processes in programming, including three phases: planning, monitoring, and optimization \cite{ma2025scaffolding,phung2025plan}. Planning concerns understanding requirements and forming an initial approach; monitoring concerns tracking progress and troubleshooting during execution (e.g., debugging); and optimization concerns improving solutions. This structured representation provides a compact profile of how the student engaged the AI during a session. The taxonomy can be customized to fit different courses and tasks. For brevity, detailed code definitions and examples are provided in the supplementary material.\footnote{\url{https://bit.ly/3O2BHDf}\label{appendix}}

Below the distribution panels, the dashboard provides a concise \textit{meta-reflective session summary} that narrates the student’s likely intent, the main difficulties encountered, and how the interaction progressed. Finally, a \textit{Detected Risks} section surfaces lightweight risk signals (e.g., overreliance on direct answers) together with brief notes that can guide instructor follow-up. This design aims to provide information necessary for pedagogical action while minimizing unnecessary access to raw conversational content.

\section{Evaluation}

We conducted a formative evaluation to assess whether MRDash provides interpretable, useful, and privacy-acceptable summaries for instructor sensemaking. The goal of this first-cycle study was to identify usability issues, perceived value, trust and privacy concerns, and concrete feedback for the next iteration.

\subsection{Participants}  

We invited six university instructors and eight student participants to take part in a formative evaluation of the prototype. This sample size aligns with prior usability research suggesting that a small number of participants can uncover most major usability issues in early-stage systems \cite{albert2022measuring}. Among the instructors, five were male and one was female, aged 32–60, and all had experience teaching programming or computing-related courses. Among the student participants, four were undergraduates and four were graduate students, six were male and two were female. All students had previously taken programming courses and reported prior experience using AI tools for programming-related tasks.

\subsection{Procedure} 

The evaluation was conducted as part of a 3-hour workshop. The prototype was deployed on a local server and populated with real-world student–AI dialogue data from an introductory programming class in 2024. The dataset covered one cohort of 22 students and 230 help-seeking sessions. For the workshop, all student identifiers (e.g., names and student IDs) were replaced with pseudonyms following standard de-identification procedures to protect student privacy.

We first introduced the project goals and the overall concept of the meta-reflective dashboard. Participants were then guided through the prototype with a short walkthrough. After the introduction, participants explored the dashboard freely for approximately 30 minutes. They could navigate the class view and open session cards to examine session summaries, prompt-type distributions, and potential risk signals. After the exploration, participants completed a feedback questionnaire. For instructors, the questionnaire included five-point Likert-scale items assessing interpretability, perceived usefulness, trust, privacy acceptability, and perceived workload reduction, as well as open-ended questions about student privacy boundaries, most/least liked features, and desired improvements for the next version. For students, the questionnaire focused on their perceptions of the session summaries, including perceived helpfulness for learning, comfort with what information is shown/shared, and willingness to receive such summaries after AI-mediated learning sessions, along with open-ended questions about what kinds of summaries feel supportive versus surveillant and what information they consider appropriate to share with instructors. Finally, we held a 60-minute semi-structured group discussion with the instructors to elicit deeper feedback that could not be captured in the questionnaire. The discussion focused on perceived classroom value, the interpretability of risk signals, and concrete scenarios in which instructors would (or would not) intervene based on the dashboard. We took detailed notes and summarized the discussion into actionable design implications for the future iteration. Detailed questionnaire items are given in supplementary material\footref{appendix}.

\begin{figure}[t]
\centering
\includegraphics[width=\textwidth]{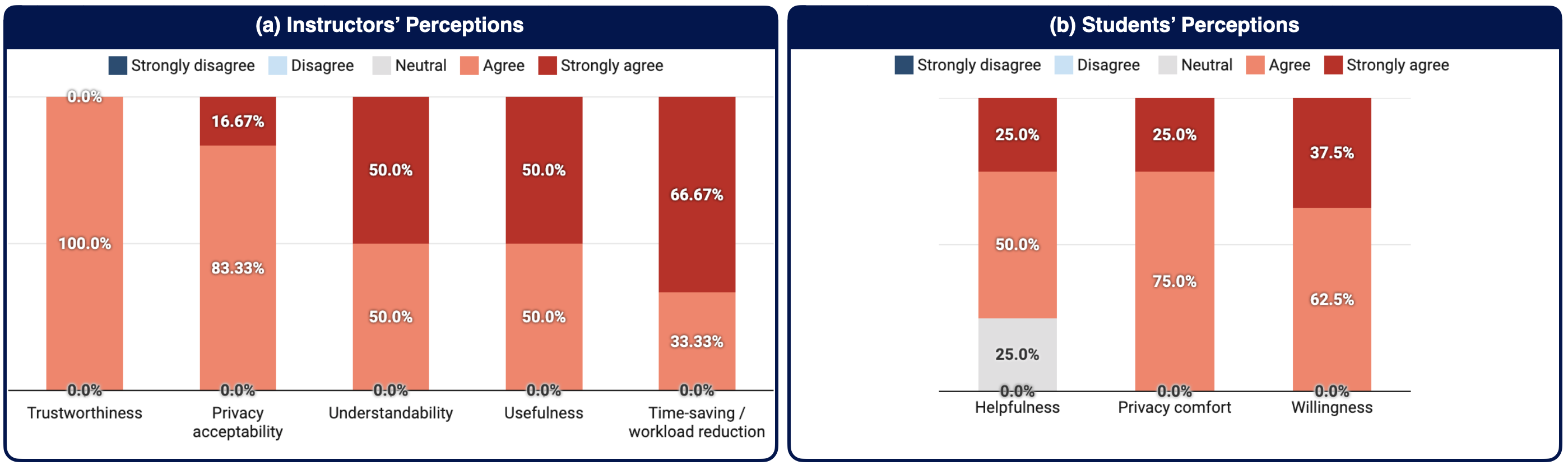}
\caption{Responses of (a) instructors’ and (b) students’ perceptions of MRDash.}
\label{evaluation}
\end{figure}

\section{Results}

\subsection{Formative Evaluation Results from Instructors}

As shown in Figure~\ref{evaluation}(a), the Likert-scale results from our formative evaluation indicate consistently positive perceptions of MRDash from instructors. Instructors (n=6) rated the dashboard as highly understandable and useful (both M=4.50; 3 \emph{Agree}, 3 \emph{Strongly agree}), and perceived strong time-saving/workload reduction benefits (M=4.67; 2 \emph{Agree}, 4 \emph{Strongly agree}). Trustworthiness was uniformly high (M=4.00; all \emph{Agree}), and privacy acceptability was also favorable (M=4.17; 5 \emph{Agree}, 1 \emph{Strongly agree}).

\subsubsection{General Impressions}
Overall, instructors responded positively to the dashboard’s high-level, structured representation of student--AI sessions. They valued being able to quickly grasp what a student was trying to do and how the interaction unfolded, without reading the full transcript. For example, one instructor noted that ``\textit{the high-level summaries can help me quickly understand students' thinking process and questioning process, while also saving time}.''

Several instructors highlighted that the prompt-type counts and distributions made sessions easier to interpret at a glance. One instructor pointed to the prompt-type frequency view, stating that ``\textit{the frequency of prompt types function helps me categorize students' questions into different strategies, which is helpful}.'' Another instructor similarly valued the session-level distributions, explaining that ``\textit{the counts and distributions of the prompt type in the Session View allow me to confirm the high-level engagement between students and AI}.''

Instructors also emphasized the practical benefit of session-level analysis for reducing workload while respecting privacy. For example, one instructor wrote that ``\textit{session analysis that shows student-AI interaction summary decreases teachers' workload while preserving student privacy}.'' Finally, instructors appreciated that the dashboard surfaced actionable signals for teaching. One instructor described that ``\textit{detected potential risks are directly useful for improving the class},'' while another noted that ``\textit{the suggestions are helpful for classroom management in real-time}.'' 

\subsubsection{Configurable Transparency and Evidence}
Open-ended responses indicated that some instructors wanted stronger evidential grounding behind the session-level summaries. In particular, instructors noted that the current prototype does not show the student’s exact prompts or the AI’s full responses, which can make it harder to judge how conclusions were reached. One instructor wrote that ``\textit{currently I cannot see the student's specific prompt and the AI's specific answer. I understand there is a tradeoff, but the system could at least provide an option to inspect details for at-risk students}.'' Another instructor proposed shifting part of this control to learners, noting that ``\textit{we can let students choose how much should be shown to teachers}.'' Together, these comments point to a need for configurable visibility---defaulting to high-level, privacy-aware summaries, while allowing controlled, case-by-case access to additional evidence when warranted.

\subsubsection{Actionable Class-Level Insights}
Instructors also expressed a clear desire for richer, more teaching-oriented analytics at the class level. For example, one instructor noted that ``\textit{In the Class View, the system should allow instructors to rank the student list by different dimensions},'' suggesting the need for customization. Another instructor emphasized the value of higher-level classroom insights, writing that ``\textit{frequently asked topics or mistakes in the class should be shown},'' and further suggested adding a longitudinal perspective by noting that ``\textit{transition of usage pattern in the time series would be useful.}'' These requests suggest that, beyond session summaries, instructors need class-level signals and configurable views that better support instructional decision-making.

%\subsubsection{Instructor Access Boundaries}
%Instructors expressed views about access boundaries. One instructor argued that teachers should be able to see students’ classroom learning behaviors to provide better support, while acknowledging privacy concerns and viewing the dashboard as a compromise: ``\textit{students’ classroom behaviors should be open to teachers... I also understand students’ desire to protect privacy, so this dashboard may be a middle-ground approach}.'' At the same time, several instructors emphasized that the system should not expose raw conversation content by default. For example, one instructor noted that ``\textit{the raw conversation logs should keep secret from instructors},'' warning that ``\textit{full monitoring may decrease the willingness of students to use AI}.'' Another instructor similarly argued that ``\textit{the original request message texts should not be shown because abstracted information would be sufficient for instruction.}'' 
%

\subsection{Formative Evaluation Results from Students}

As shown in Figure~\ref{evaluation}(b), students also responded positively. For \textit{Helpfulness}, students rated the summaries as beneficial for learning (Mean = 4.00; 4 \textit{Agree}, 2 \textit{Strongly agree}). For \textit{Privacy comfort}, students reported being comfortable with what information is shown or shared (Mean = 4.25; 6 \textit{Agree}, 2 \textit{Strongly agree}). For \textit{Willingness}, students indicated that they would be willing to receive such summaries after AI-mediated sessions (Mean = 4.38; 5 \textit{Agree}, 3 \textit{Strongly agree}). These distributions suggest that students generally accepted the summaries as a low-friction reflection aid, while also motivating careful attention to privacy framing and meaning of risk indicators.

\subsubsection{Perceived Support vs. Monitoring}
Open-ended responses suggested that students generally viewed the dashboard as learning support rather than a means of enforcement. Several students emphasized that the summaries could help instructors better understand their learning situation and provide more appropriate support. For example, one student noted, ``\textit{It feels supportive because it helps the instructor understand my situation and help me fill knowledge gaps in my learning},'' while another highlighted that the dashboard focuses on learning-relevant signals rather than private content: ``\textit{it is good, since it only shows my learning approach rather than privacy-sensitive chat logs}.''

A few described a mixed feeling of being both supported and monitored, but stressed that this was acceptable as long as the data were used strictly for educational purposes: ``\textit{I feel both helped and monitored, but I do not really mind being monitored as long as the data are not used beyond education}.'' Another student reframed the experience as closer to self-supervision than external surveillance, explaining that the signals functioned as a reminder to use AI for understanding rather than cheating: ``\textit{It feels less like being monitored and more like supervising---AI should be used to support learning and understanding, not for cheating}.'' 

\subsubsection{Student Information-sharing Boundaries}
Students also drew boundaries around what information should be shared with instructors. In general, they were comfortable sharing high-level, learning-relevant summaries that help instructors understand their learning status and provide support. For example, one student noted that ``\textit{an overall summary and synthesis of my questions can be shared with teachers}.'' Students also specified which signals they considered appropriate to share. One student suggested that instructors could see learning progress, difficult concepts, and the reasoning process toward a solution, explaining that ``\textit{these can help teachers understand my learning situation and provide precise guidance}.'' Another student similarly noted that it is acceptable to share the knowledge concepts they asked about as long as the information is abstracted: ``\textit{What knowledge concepts I asked about can be shared, but it should be abstracted; details should not be shared}.''

At the same time, students expressed discomfort with sharing raw or overly detailed records. One student stated that they would prefer not to share “the specific conversation content or how many questions I asked,” expressing concern that instructors might infer they had asked too many questions. Another student drew a firm line around personal content and exploratory thinking, stating that ``\textit{immature ideas or incorrect hypotheses generated during the conversation should also remain private}.'' Finally, one student stressed that acceptability depends on purpose, warning that ``\textit{while the summaries can reflect learning mastery by showing weaknesses through the questions asked, using them to indicate plagiarism in ways that affect grades could create opposition between teachers and students and reduce students' willingness to use AI tools}.''

\subsubsection{Benefits Beyond Instructors}
Although the primary audience is instructors, students suggested that the summary can also benefit their own learning. Several students noted that a structured summary could surface issues they might otherwise overlook. For example, one student explained that ``\textit{when I ask AI for help, I often do not know what I am lacking, and this would help me notice points I did not pay attention to before}.'' Another student valued a more objective lens for self-regulation, stating that ``\textit{it could help me understand from a third-party, objective view whether my learning behaviors are appropriate and how to learn more efficiently, and avoid being wrong without realizing it}.''

Students also emphasized that session summaries could support reflection by clarifying what they have mastered and what remains uncertain. One student noted that ``\textit{it helps reveal what knowledge I have mastered and which parts I understand only vaguely},'' while another added that reviewing the summary can quickly expose flawed reasoning and prevent repeated mistakes in similar tasks: ``\textit{I can immediately see my incorrect reasoning and avoid making the same mistake next time}.''

%
%\section{Discussion}

\section{Conclusion}

This paper presents a meta-reflective dashboard that summarizes individual student–AI interactions into concise, interpretable reports for instructors. Rather than exposing full chat logs, the dashboard surfaces high-level interaction patterns, AI usage behaviors, and potential pedagogical or integrity risks, accompanied by brief, actionable follow-up suggestions. Our goal is to provide instructors with a low-effort way to make student–AI help-seeking visible and interpretable while respecting privacy constraints.

There are limitations for this paper. First, we report a first-cycle, formative evaluation of a prototype dashboard. Findings are based on a small sample and therefore primarily reflect perceived usefulness, interpretability, trust, and privacy acceptability rather than demonstrated impacts on teaching practice. The prototype was evaluated with data from a single introductory programming course, so generalizability to other courses, domains, institutions, and assessment contexts remains uncertain. We also did not observe longitudinal, in-situ use, including potential behavioral changes or chilling effects on help-seeking. Finally, because the dashboard relies on AI-generated summaries and risk signals, outputs may be incomplete or inaccurate and may vary under different models. Further work is needed to validate signal quality, examine instructor interpretation and fairness, and refine governance mechanisms such as configurable transparency and consent. Building on these results, future work will iterate on the dashboard design with instructors and students and evaluate it in authentic classroom deployments to examine how instructors use the summaries and signals in practice, and how such support affects teaching decision-making, student help-seeking, and perceptions of privacy and accountability.

%Limitation :我们只用了编程课，然后category也可以用

%
% the environments 'definition', 'lemma', 'proposition', 'corollary',
% 'remark', and 'example' are defined in the LLNCS documentclass as well.
%


\bibliographystyle{splncs04}
\bibliography{bib}

\end{document}